\documentclass[final]{aipproc}
\usepackage{amsmath,amssymb,wasysym}
\usepackage{graphicx}

\layoutstyle{8x11single}

\begin{document}

\title[Towards photometry pipeline of the Indonesian space surveillance system]{Towards photometry pipeline of\\ the Indonesian space surveillance system}


\author{Rhorom Priyatikanto}{
  address={Space Science Center, National Institute of Aeronautics and Space (LAPAN)},
  altaddress={Astronomy Program, Faculty of Mathematics and Natural Sciences, Institut Teknologi Bandung},
  email={rhorom.priyatikanto@lapan.go.id}
}

\author{Bahar Religia}{
  address={Astronomy Program, Faculty of Mathematics and Natural Sciences, Institut Teknologi Bandung}
}

\author{Abdul Rachman}{
  address={Space Science Center, National Institute of Aeronautics and Space (LAPAN)}
}

\author{Tiar Dani}{
  address={Space Science Center, National Institute of Aeronautics and Space (LAPAN)}
}

\begin{abstract}
Optical observation through sub-meter telescope equipped with CCD camera becomes alternative method for increasing orbital debris detection and surveillance. This observational mode is expected to eye medium-sized objects in higher orbits (e.g. MEO, GTO, GSO \& GEO), beyond the reach of usual radar system. However, such observation of fast-moving objects demands special treatment and analysis technique. In this study, we performed photometric analysis of the satellite track images photographed using rehabilitated Schmidt Bima Sakti telescope in Bosscha Observatory. The Hough transformation was implemented to automatically detect linear streak from the images. From this analysis and comparison to USSPACECOM catalog, two satellites were identified and associated with inactive Thuraya-3 satellite and Satcom-3 debris which are located at geostationary orbit. Further aperture photometry analysis revealed the periodicity of tumbling Satcom-3 debris. In the near future, it is not impossible to apply similar scheme to establish an analysis pipeline for optical space surveillance system hosted in Indonesia.
\end{abstract}

\maketitle

\section{Introduction}
Human civilization has entered space age for about fifty years, filled the Earth surrounding space with man-made orbital objects. According to the catalog released by US Space Surveillance Network, there are more than 17000 cataloged objects in orbit, mostly categorized as fragmented debris ($\gtrsim9000$ objects) that move reinlessly \citep{odq182}. High concentration of orbital objects occupy Low Earth Orbit (LEO, $\sim800$ km above sea level) and Geostationary Equatorial Orbit (GEO, $\sim36000$ km above equator). Even though various process trigger natural re-entry of objects in low orbit, fragmentation caused by two-body collision, on the other hand, increases space debris population by a significant amount. Intended firing of Fengyun 1C and accidental collision between Cosmos 2251 and Iridium 33 drastically rise the potential threat of space debris to the active satellites \citep{odq182}. Continuous observation and monitoring of orbital objects become important tasks to ensure the safety of space activities.

Orbital and space debris observation can be conducted in radio wavelength using Radar technology (e.g. Space Surveillance Network) and in optical-infrared window using fast-response telescope. Optical observation provide some advantages compared to Radar-based observation. Firstly, optical observation relies on the Solar radiation reflected by orbital objects to the observer so that it can detect object located in higher orbit since observed flux drops quadratically ($F\propto1/h^4$ for Radar observation). The actual detection range depends on the size and albedo of the object. Besides, optical observation could serve supplementary data since Radar observation commonly fails to detect orbital object with low reflectivity in radio wavelength \citep{potter1995}.

Optical observation of space debris utilize sub-meter telescope with fast-response and high agility \citep[e.g.][]{koltz2008} though the detection limit of small telescope is brighter. High sensitivity Charge-Coupled Device (CCD) is used as the standard detector for both astrometry and photometry purposes. The telescope is operated in various mode to detect fast-moving objects.
\begin{itemize}
\item\textbf{Trailing mode} where the telescope is pointed to celestial object or the predicted location of any orbital object and tracks the celestial object with sidereal angular speed ($\omega\approx15'$ per minute).
\item\textbf{Tracking mode} where the system automatically identifies and follows the movement of orbital object in celestial plane \citep[e.g.][]{koltz2008}. Portrait of a point-like object among star trails is the observation result of this mode.
\item\textbf{Hybrid mode.} Instead of tracking the orbital object physically, special digital processing can be done to track the object. \emph{Time delay integration} (TDI) is an example of technique for hybrid mode observation \citep{seitzer2004}.
\end{itemize}

As the continuation of space surveillance program \citep{rachman2009}, National Institute of Aeronautics and Space (LAPAN) started the development of orbital object/debris observation in optical window, especially to monitor objects in low orbit with high probability of re-entry \citep{dani2014}. As the main part of the system, astrometry and photometry analyses scheme are established in order to determine the orbital parameter \citep{rachman2014} and photometric properties of the observed objects \citep[e.g.][]{yanagisawa2012,seo2013}.

This work focuses on the photometric analysis scheme of the images of orbital object observed in trailing mode. The purposes of the analysis are to identify the streak pattern of fast-moving object  and to extract the flux as a function of time. Section 2 discusses the analysis scheme that consists of Hough transformation \citep{hough1962} as the backbone tool to automatically detect the satellite trail and aperture photometry to extract the flux. Section 3 presents the observational data obtained in Bosscha Observatory and the implementation of analysis scheme to the data. Discussion and foresight are presented in Section 4.

\section{Method}

\begin{figure}
\centering
\includegraphics[width=0.8\textwidth]{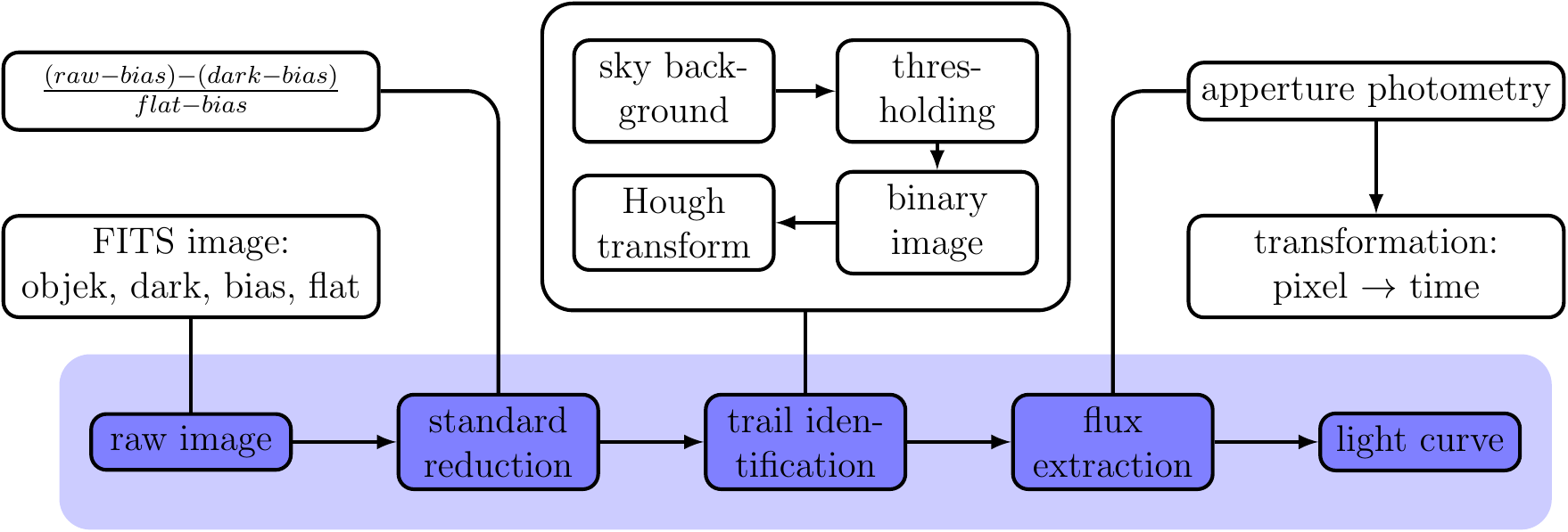}
\caption{Schematic diagram of the photometry pipeline to extract the flux of orbital object.}
\label{scheme}
\end{figure}

The proposed analysis pipeline consists of three main parts (Figure \ref{scheme}): (1) trail identification, (2) flux extraction, and (3) light curve construction. The identification of fast-moving object on the frame relies on Hough transform \citep{hough1962} as standard tool for line detection \citep{storkey2004}. This process transforms the position any pixel in $2D$ image ($x,y$) into sinusoidal curve in Hough space ($\theta,\rho$). In this way, a line crossing any point is parameterized by its perpendicular distance from the origin ($\rho$) and the orientation angle ($\theta$)\footnote{The reader may visit \url{www.storkey.org/hough.html} for visual demonstration of Hough transform.}.
\begin{equation}
(x,y) \longrightarrow (\theta,\rho);\quad \rho=\rho_m\cos\left(\theta-\theta_m\right),
\end{equation}
with $\rho_m=\sqrt{x^2+y^2}$, $\theta_m=\tan^{-1}\left(\frac{y}{x}\right)$ and $\theta=[0,\pi]$. A linear pattern in the frame will be the intersection of many sinusoidal curves in the Hough space. In practice, both $\rho$ and $\theta$ are discretized for counting so that a prominent line/streak in the image corresponds to ($\theta,\rho$) with the highest count.

Hough transform could be applied to gray-scale images, but it is more convenient to implement it to binary images since satellite trail usually dimmer than stars. Binary images are created through thresholding that uses sky background ($c_{sky}$) as the reference to determine thresholding value, $c_{thres}$. The value of $c_{sky}$ is calculated using $3\sigma$ clipping \citep[e.g.][]{bertin1996} and $c_{thres}\approx1.2c_{sky}$ is appropriate for satellite images with relatively low $S/N$. However, wide field image such as photographed using telephoto lens \citep{dani2014} may displays inhomogeneous sky background due to light pollution or twilight. To deal with this issue, $c_{sky}$ is calculated from sub-images to model the variation of sky background value.

After Hough transformation, ($\theta,\rho$) with maximum count is selected to represent single satellite trail. Flux of the object along the trail is extracted using aperture photometry as the ideal approach for relatively isolated object without any assumption of the point-spread function \citep{howell2000}. The observed pixel counts are accumulated of a box with typical size of 8 pixel that corresponds to $\sim3\cdot FWHM$. The obtained counts are normalized with the area of the box and subtracted by surrounding sky background value. Flux as a function of pixel position is the transformed into light curve assuming that the length of satellite trail correspond to the total exposure time.

\section{Data and Analysis}

\begin{figure}
\centering
\framebox{\includegraphics[width=4.5cm]{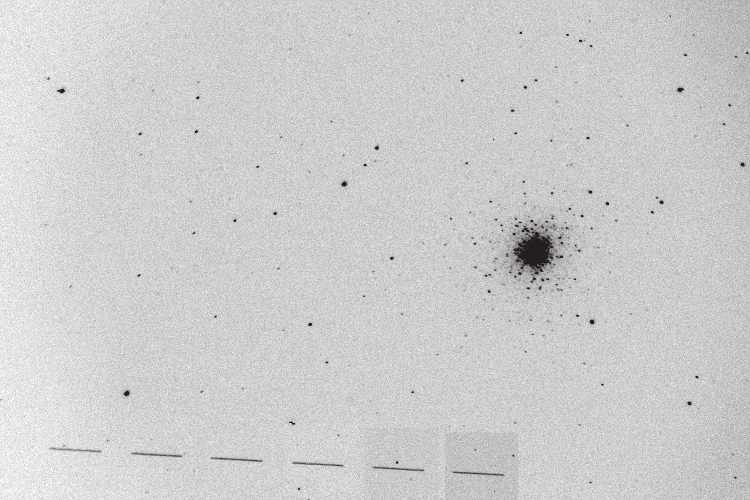}}\hspace{5pt}
\framebox{\includegraphics[width=4.5cm]{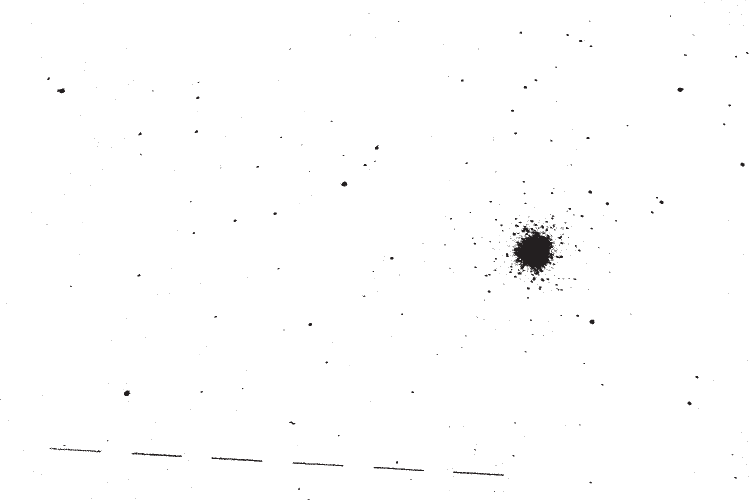}}\hspace{5pt}
\framebox{\includegraphics[width=4.5cm]{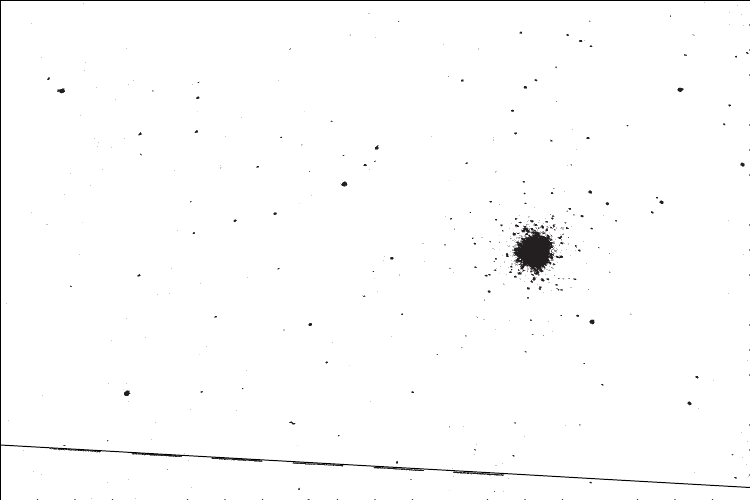}}
\caption{Composite image of six consecutive frames, with 10 s exposure time and 5 s interval between two exposure, displays Thuraya-3 satellite trail (left). Thresholding process results binary image (center) with clear satellite track. Trail is identified (right) according to statistics in Hough space.}
\label{thuraya3}
\end{figure}

\begin{figure}
\centering
\includegraphics[height=4.5cm]{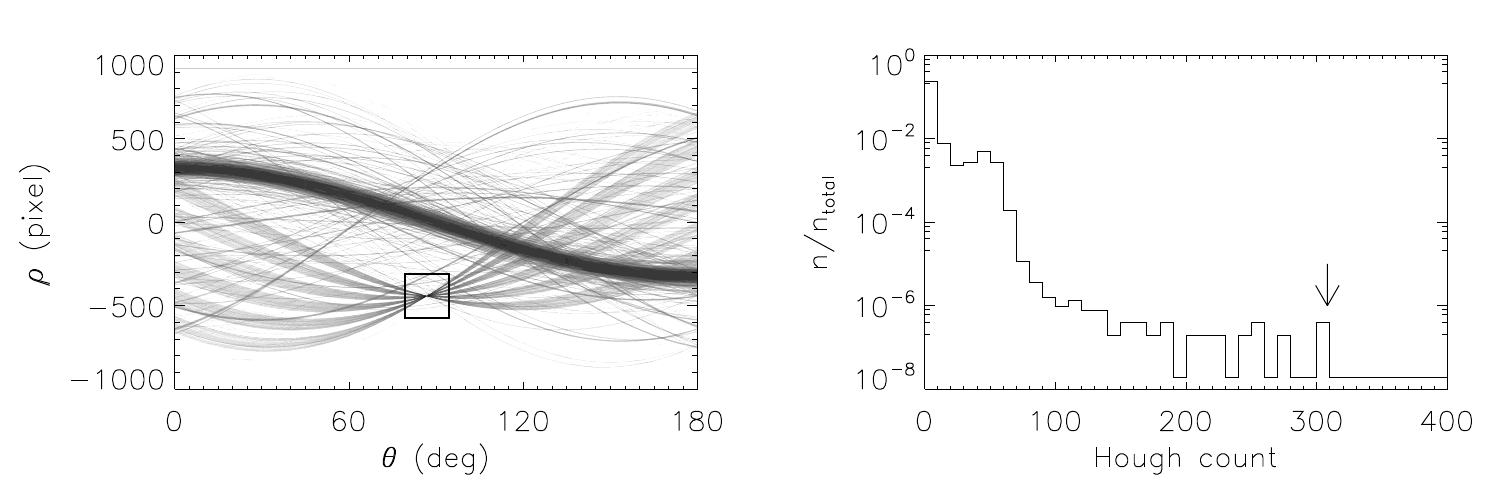}
\caption{Transformation results plotted on the Hough space (left) shows prominent intersection/node marked with a rectangle. Every set of ($\theta,\rho$) has its count. The density distribution of the counts follows Poisson distribution with outliers that correspond to the linear pattern in the image (right).}
\label{hough}
\end{figure}

Observational data are obtained by Bahar Religia on June 22nd 2014 using rehabilitated Schmidt Bima Sakti telescope ($D=51$ cm, $f=127$ cm) in Bosscha Observatory \citep{akbar2013}. CCD camera with $1530\times1020$ pixels (9 mikron each) are mounted on the telescope, provides moderately wide field of view of $0.62^{\circ}\times0.24^{\circ}$ with spacial sampling of 1.47''/pixel. The instrument is also equipped with Bessel $BVRI$ filters. Raw images went through standard reduction process invoking bias, dark, and flat images. Image processing software IRIS\footnote{\url{www.astrosurf.com/buil/us/iris/iris.htm}} is used for this purpose. After reduction, images of satellite trail with signal to noise ration $S/N\sim3$ are obtained for the following analysis.

During multicolor observation of globular cluster NGC 7089 ($\alpha_{2000}=21^{\text{h }} 33^{\text{m }} 27.02^{\text{s }}$, $\delta_{2000}=-0^{\circ} 49' 23.7''$), two satellites are crossing the field of view. Those two satellites are identified as Satcom-3 Deb (crossing at 20.38 UT) and Thuraya-3 (crossing at 21.01 UT) according to updated Two-line Element (TLE) from \url{www.celestrak.com}. Cataloged orbital parameter of these objects are summarize in Table \ref{orbit}.

\begin{table}[ht]
\centering
\caption{Orbital parameters of the satellites: period ($P$), semi-major axis ($a$), eccentricity ($e$), inclination ($i$), and rightascension of ascending node ($\Omega$). These parameters are derived from the current TLE.}
\label{orbit}
\begin{tabular}{p{3cm}p{1.5cm}p{1.5cm}p{1.5cm}p{1.5cm}p{1.5cm}}
\hline
& \multicolumn{5}{c}{Orbital parameter}\\
Satellite & $P$ [hours] & $a$ [km] & $e$ & $i$ [deg] & $\Omega$ [deg]\\
\hline
Thuraya-3 & 23.928 & 42,313 & 0.00 & 4.48 & 345.24\\
Satcom-3 Deb & 12.924 & 28,063 & 0.47& 8.89 & 106.76\\
\hline
\end{tabular}
\end{table}

\begin{figure}
\centering
\includegraphics[height=4.5cm]{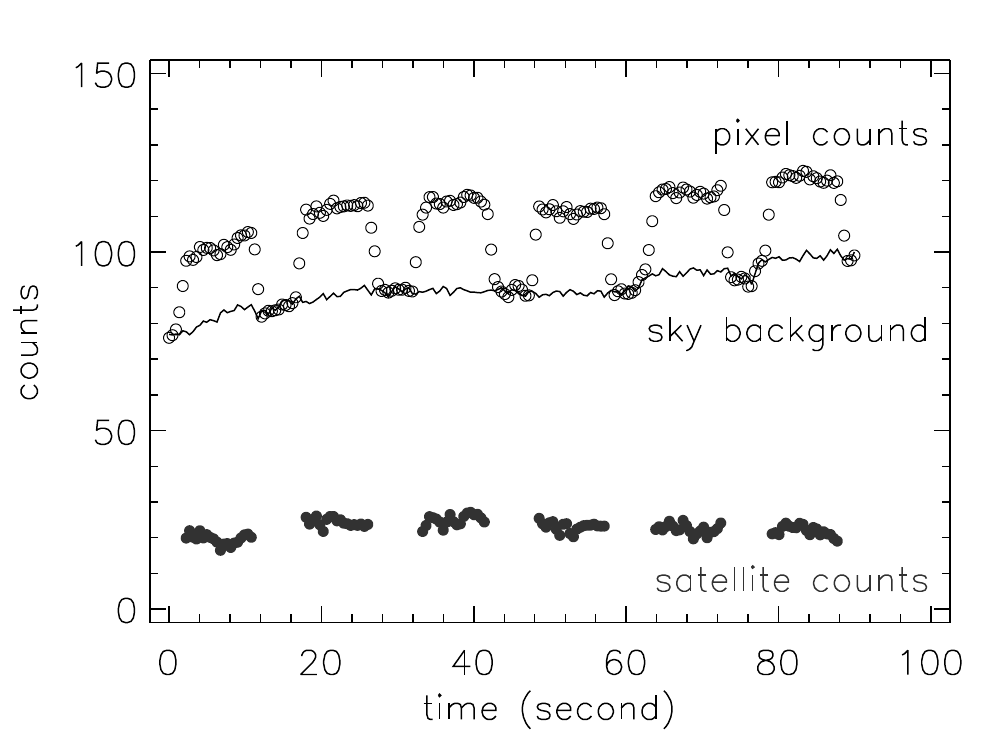}\hspace{5pt}
\includegraphics[height=4.5cm]{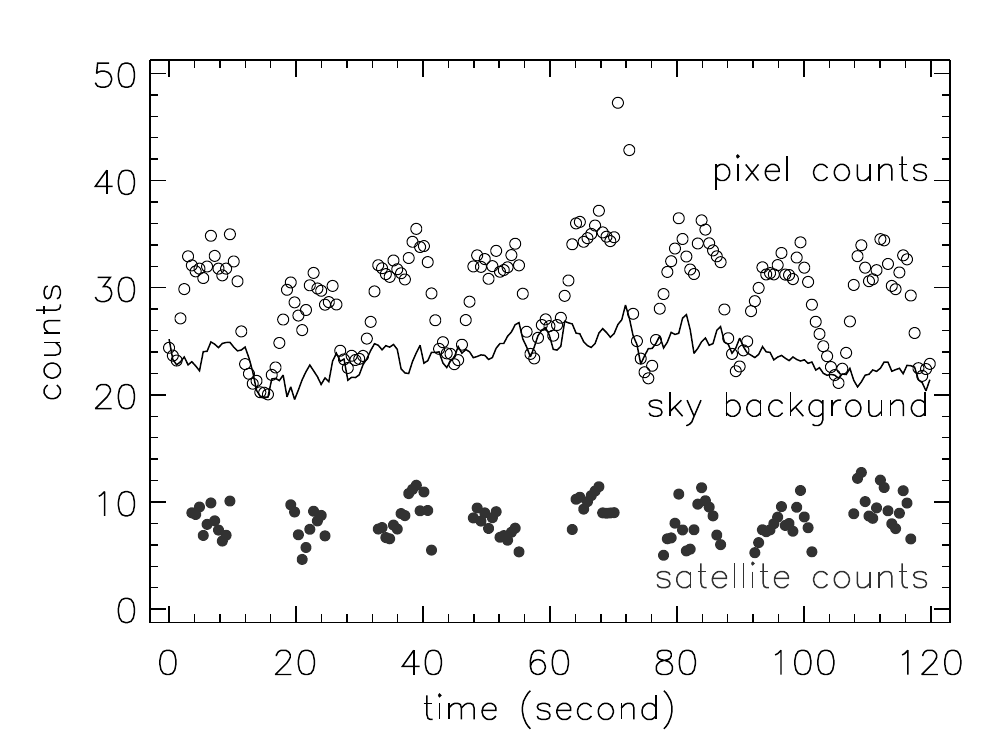}
\caption{Light curve of Thuraya-3 (left) and Satcom-3 Deb (right) as obtained from aperture photometry along the satellites' path. Sudden drops of the curve correspond to non-exposure phase (interval between two exposure).}
\label{lightcurve}
\end{figure}

Figure \ref{thuraya3} displays the detection process of Thuraya-3 communication satellite that crosses the field of view, to the south of NGC 7089. In the single image, Thuraya-3 (and also Satcom-3 Deb) creates relatively short trail ($\sim2.5'$) which is hard to detect using Hough transform because the trail is overwhelmed by the prominent cluster that spans over 10'. The detection works well to the composite image consisting half-degree-length satellite trail.

Aperture photometry of the object is conducted along the identified trail/track. The flux of Thuraya-3 and Satcom-3 Deb over time is plotted in Figure \ref{lightcurve}. The light curve of Thuraya-3 does not exhibit a significant variation, while the Satcom-3 Deb clearly shows and eratic variation with amplitude of abour 30\%. This variation of period $\sim3$ seconds indicates that inactive Satcom-3 Deb tumbles about its center of mass. However, low $S/N$ image makes the proper period determination can not be conducted.

\section{Discussion}
The photometric observation of orbital objects is essential task to determine the physical properties such as the size, albedo, and attitude \citep{potter1995,seo2013}. This work can be considered as an embryo of more comprehensive analysis pipeline that automatically detect any orbital objects and extract informations from the observational data. There are several works wait to be accomplished in near future, such as the integration to astrometry pipeline, robust modeling of sky brightness, absolute magnitude measurement, and also periodicity analysis of the light curve.

This work deals with very limited cases where the existence of satellite trails are visually detected by the observer. Automation of this process requires statistical evaluation of the output of Hough transform and becomes essential if the observational data is supplied through all-sky survey mission. Besides, wide field observational data obtained by Rachman et al. \citep{rachman2014,dani2014} challenges the analysis pipeline to extract the photometric data. Different characteristics of the data from CMOS camera demands a bit different treatment compared to CCD-based data.

\bibliographystyle{aipproc}
\bibliography{paper}

\end{document}